\documentclass[10pt,conference]{IEEEtran}
\usepackage{fancyvrb}
\usepackage{graphicx}
\usepackage{relsize}
\usepackage{booktabs}
\usepackage{url}
\usepackage{tabularx,ragged2e}
\usepackage[english]{babel}
\usepackage{verbatimbox}
\usepackage{cite}
\usepackage{amsmath,amssymb,amsfonts}
\usepackage{algorithmic}
\usepackage{graphicx}
\usepackage{textcomp}
\usepackage{xcolor}
\usepackage{multirow}

\newcolumntype{C}{>{\Centering\arraybackslash}X}
\def\BibTeX{{\rm B\kern-.05em{\sc i\kern-.025em b}\kern-.08em
    T\kern-.1667em\lower.7ex\hbox{E}\kern-.125emX}}
\usepackage{pifont}% http://ctan.org/pkg/pifont
\newcommand{\cmark}{\ding{51}}%
\newcommand{\xmark}{\ding{55}}%
\begin{document}
\title{PR-SZZ: How pull requests can support the tracing of defects in software repositories}


\author{
\IEEEauthorblockN{Peter Bludau}
\IEEEauthorblockA{\textit{fortiss GmbH - Research Institute of} \\ 
\textit{the Free State of Bavaria} \\
Munich, Germany \\
bludau@fortiss.org \\[-3ex]}
\and
\IEEEauthorblockN{Alexander Pretschner}
\IEEEauthorblockA{\textit{Technical University of Munich} \\
Munich, Germany \\
alexander.pretschner@tum.de}
}

\IEEEaftertitletext{\vspace{-140pt}}
\maketitle
\begin{abstract}
  The SZZ algorithm represents a standard way to identify bug fixing commits as well as inducing counterparts.
  It forms the basis for data sets used in numerous empirical studies. 
  Since its creation, multiple extensions have been proposed to enhance its performance.
  For historical reasons, related work relies on commit messages to map bug tickets to possibly related code with no additional data used to trace inducing commits from these fixes.
  Therefore, we present an updated version of SZZ utilizing pull requests, which are widely adopted today.
  We evaluate our approach in comparison to existing SZZ variants by conducting experiments and analyzing the usage of pull requests, inner commits, and merge strategies. We base our results on 6 open-source projects with more than 50k commits and 35k pull requests.
  With respect to bug fixing commits, 
  on average 18\% of bug tickets  
  can be additionally mapped to a fixing commit,
  resulting in an overall F-score of 0.75, an improvement of 40 percentage points.
  By selecting an inducing commit, we manage to reduce the false-positives and increase precision by on average 16 percentage points in comparison to existing approaches.
\end{abstract}



%!TEX root = main.tex
%%%%%%%%%%%%%%%%%%%%%%

\section{Introduction}

% Situation
Bug introducing changes in software repositories are an important asset to empirical software engineering research. 
 Many studies rely on defect data sets to either 
 analyze defects and investigate the characteristics of bug inducing activities \cite{Eyolfson2011, Kononenko2015, Chen2019c} or
 evaluate defect prediction approaches \cite{Kamei2013a, Catolino2017a, Pascarella2019}.
The standard algorithm used to extract information about bug fixing and inducing commits from the version control system (VCS) is called SZZ,
 proposed by \'{S}liwersky, Zimmermann and Zeller \cite{Jaceksliwerski2005} in 2005.
 The algorithm can be described as a two-step process.
 The first part maps commits from the version control system to resolved tickets from an issue tracking system (ITS).
 Commits that contain specific information about resolved bug labeled tickets 
 are identified as "bug fixing commits".
 The second part searches for changes that potentially introduced the bug.
 The algorithm finds modified lines for the fixing commits and traces them in the VCS history to reveal a version that last changed these lines.
 This reveals the "bug inducing commits".
 Especially, the second part of the algorithm was subject to many studies \cite{Kim2006, Williams2008, Davies2014, Costa, Neto, Neto2019} to cope with inaccuracies of the original version. These studies propose additions to the tracing
 of inducing commits by filtering false-positives (commits that are falsely labeled bug inducing).
 As shown in existing studies, there are still open issues regarding mapping a bug ticket to a fixing commit (e.g., \cite{Bird2009, Bissyande2013, Herbold2019}) and finding the correct inducing commits (e.g., \cite{Herzig2013, Mills2018, Costa}).

% Complication
The first part of the algorithm relies on matching a commit to a bug ticket and thereby reveals the bug fixing commit. 
Multiple problems may arise in this process. 
Due to commits that do not mention the related issue number or do not have well-defined commit messages some bug fixing commits cannot be mapped to their respective bug ticket.
These are missing in the resulting data set which may lead to highly imbalanced bug data sets containing a fraction of all bug fixing commits (false-negatives).
Moreover, commits can be wrongly selected as bug fixing commits for a bug ticket (false-positives).
Among other circumstances, this is caused by links between bug tickets and commits that are wrongly inferred by the algorithm due to incorrect commit message parsing. 
This has also an impact on the second step of the algorithm since wrong bug fixing commits are used to trace inducing commits.

%Complication II
With already proposed enhancements to the second part of the algorithm, some open issues remain.
Bug fixes that only add new code, for instance adding a missing null-check to fix a null pointer exception, are not found by the algorithm because they cannot be blamed onto existing changes. 
If this code is added together with other changes in a commit, correct but unrelated code is possibly labeled as bug inducing.
Moreover, since large fix commits are potentially reporting many inducing commits, they also tend to introduce many false-positives \cite{Costa}.
%Complication III
With too many commits blamed as bug inducing for such large fixing commits and on average 43\% of defective commits being indeed only partially defective (containing not only defective files)
\cite{Pascarella2019}, results of SZZ may be too coarse.
To the best of our knowledge existing SZZ variants propose additions to reduce false-positive bug inducing commits, but use the same information from VCS history and issue tracking systems
proposed more than 15 years ago to find fixing and inducing changes for a certain bug ticket on the commit level.

% Resolution
This study is meant to address aforementioned issues by proposing an extension to existing SZZ variants utilizing additional data from open-source development.
In modern software development, integrated information is available that more thoroughly describes the relationships between existing code, bug tickets, and contributed changes.
For instance, on code management platforms (CMP), e.g., GitHub, developers can propose changes in a pull-based development workflow \cite{GPD14} that incorporates reviews and discussions regarding new and changed code. 
Thereby, actions taken to fix a bug are collected in pull requests.
Such pull requests are already used as sources for additional bug reports and patches \cite{franco2017, garcia2020, waan2017}. 
However, we are not aware of any existing research that uses data from pull requests and inner feature development to support the SZZ algorithm in a systematic way to
retrieve additional fixing and more fine-grained inducing commits.
Like the original algorithm, our work can be divided into two parts.
First, we extract links between pull requests, ITS tickets and commits from various platforms 
to find fixing commits even when the issue is not directly referenced in the commit. 
Secondly, to find inducing commits our approach analyses the development activities inside a pull request to filter unrelated changes and propose more fine-grained results.
We conclude our work by creating data sets not only on commit, but also on file and method level, since data sets on a commit level can be too coarse to filter partially inducing commits.
 
% Contribution
Our contributions can be summarized as follows:
\begin{itemize}
    \item We design a new version of the SZZ algorithm that relies on the usage of pull requests and associated data. Thereby, we aim at finding more correct bug fixing commits from the bug tickets, as well as, retrieving more correct sets of bug inducing commits.
    \item We contribute to both parts of the algorithm.
    Searching for fixing commits, our approach increases the number of found fixes while obtaining high performance values on a manually validated data set.
    Tracing inducing commits using our approach, the precision is increased for all investigated projects.
    \item We publish defect data sets containing bug fixing and inducing commits for 6 well-known open-source projects on commit, file, and method level to enable further research.
\end{itemize}

%%%%%%%%%%%%%%%%%%%%%%%
\section{Background} \label{BackgroundRefs}
In the first part of this section, we describe the inner functioning of the original SZZ approach as well as summarize the adjustments and improvements already proposed by other studies.
Our approach is based on the incorporation of data generated in open-source development workflows.
Therefore, we discuss modern software development activities and processes in the second part.

\subsection{The SZZ Algorithm}

To refer to each of the following presented variants of the SZZ algorithm we use the short notation introduced by Costa et al. \cite{Costa} and extended by Rosa et al.\cite{Rosa}.
Table \ref{tab:table_szz_variants} shows the SZZ variants proposed in existing literature.
 
 The original SZZ algorithm (B-SZZ) \cite{Jaceksliwerski2005} uses an ITS, e.g., Jira, and 
 a VCS, e.g., git, 
 to find bug fixing and inducing commits in the commit history.
 B-SZZ works in two phases with the first part searching the fix commit for a given bug ticket.
 Each bug ticket from the ITS is collected. Matching fixing commits are searched by applying two confidence checks to the commits.
 The first check verifies the syntactic conformity.
 Parsing the commit message it looks for a link to a bug number and/or keyword (e.g., 'fix').
 The second check verifies the semantic association by comparing characteristics of the syntactically linked bug ticket and commit.
 For example, the description of the bug should be contained in the commit message or
 the assignee to the bug ticket should also be the one that authored the commit.
Depending on the number of met criteria, a commit is either discarded or added to potentially fixing commits. 
The newest potentially fixing commit is selected as the fixing commit.
In the second part, B-SZZ traces bug inducing commits.
The algorithm computes the differences for each fixing commit compared to its direct parent.
Thus, the lines changed in the fixing commit are obtained.
The last commit from the VCS history that changed said lines is retrieved by using the VCS \textit{annotate} feature. 
All found commits form the list of bug-inducing suspects.
To reduce false-positives, the algorithm rejects changes that are commited to the VCS after the bug ticket was created.

\begin{table}
\caption{Proposed variants of the SZZ algorithm in literature.}
%\begin{minipage}{\columnwidth}
\begin{center}
\begin{tabularx}{\columnwidth}{ll}
\toprule
    Variant & Authors \\%& Short Description\\
    \hline
    B-SZZ   
        & {S}liwersky et al. \cite{Jaceksliwerski2005} in 2005 
        \\
    AG-SZZ  
        & Kim et al. \cite{Kim2006} in 2006  
        \\ 
    DJ-SZZ  
        & Williams and Spacco \cite{Williams2008} in 2008 
        \\
    L-SZZ \& R-SZZ
        & Davies et al. \cite{Davies2014} in 2014
        \\
    MA-SZZ  
        & Da Costa et. al. \cite{Costa} in 2017   
        \\
     A-SZZ 
        & Sahal and Tosun. \cite{Sahal2018} in 2018  
        \\
    RA-SZZ 
        & Neto et. al. \cite{Neto, Neto2019} in 2018, 2019   
        \\
\bottomrule
\end{tabularx}
\label{tab:table_szz_variants}
\end{center}
%\end{minipage}
\end{table}

The variants of B-SZZ shown in Table \ref{tab:table_szz_variants} do focus on improving the second part of the algorithm.
 Kim et al. \cite{Kim2006} found that cosmetic changes to the code can have an effect on found bug inducing commits. 
 The proposed AG-SZZ uses the annotation graph of the VCS to trace changes in the version history while ignoring changes to comments and white spaces.
 They further propose to discard changes to code style and formatting and remove outlier fix revisions, 
 commits that change disproportionately many files, 
 to reduce the false-positive rate.
Willams and Spacco \cite{Williams2008} pointed out that the results of the annotation graph described by Kim et al. \cite{Kim2006} can be imprecise considering large change hunks in a file. 
They propose DJ-SZZ, a line-number mapping approach, to track each uniquely changed line across the revisions in the VCS history.
In addition, they extend the notion that non-semantic changes should be neglected by the algorithm. 
They use DiffJ \cite{Pace2007} to ignore changes to lines containing comments, imports, or method signatures.
As stated by Da Costa et al. \cite{Costa}, filtering meta-changes
\textendash~changes that are not related to source code and cannot introduce bugs \textendash~results in more accurate bug inducing commits.
Their version, denoted as MA-SZZ, discards git meta changes, for instance merge commits
and file meta changes, such as file property or permission updates.
Lines that are only added to a commit without an associated removed line cannot be tracked by SZZ because there is no line in the VCS history responsible for its introduction. 
Sahal and Tosun \cite{Sahal2018} describe another variation that takes the surrounding code block into account.
They trace the changes for all lines from the code block with an addition to find potentially inducing commits.
Neto et al. \cite{Neto, Neto2019} introduced a refactoring aware version of the SZZ algorithm (RA-SZZ).
They propose to restrict changed lines from the fixing commits by ignoring changes that are flagged as refactorings.
First, they apply the tool RefDiff \cite{Silva2017} to filter refactorings in Java code. 
Later, they enhance their previous results by replacing RefDiff with RefactoringMiner \cite{Tsantalis:ICSE:2018:RefactoringMiner}.
 
Since SZZ versions select multiple commits as bug inducing, Davies et al. \cite{Davies2014} proposed two adjusted selection mechanisms to reduce the number of bug inducing suspects. Their first mechanism, L-SZZ, considers only the largest commit among all suspected commits as bug inducing. The second one (R-SZZ) selects the most recent commit from the list of suspects as the defective commit.

%%%%%%%%%%%%%%%%%%%%%%%

\subsection{Collaborative Software Development}

In order to support collaborative software development, pull-based development \cite{GPD14} is well established.
It promotes a synchronisation mechanisms to incorporate change proposals from a developer into a central code repository by asking for permission to pull changes into it.
Typically, a pull request is created for a single feature development or bug fix and therefore is valid in its own context. 
It includes all information that is needed to discuss, review and resolve the proposed changes.
The decision if a pull request is either accepted or rejected is taken by the core developers of a repository \cite{Gousios2015}. 
They resolve due discussions and assign reviewers to respective pull requests. 
The available information on code changes paired with the remarks expressed by contributors, reviewers and core developers represents a valuable asset to research on the code integration process.
When a pull request is accepted, different projects use various merge strategies to integrate the changes \cite{7887704}.
For instance, changes can be merged creating a merge commit that integrates two development strands into one, without changing the actual development history. 
Contrary, they can also be rebased, rewriting the VCS history and generating new commits.
Squashing pull request when integrating changes is a common technique to reduce the number of commits created in a pull request to the VCS.
However, information about development activities is lost in the process.

%%%%%%%%%%%%%%%%%%%%%%%

\section{A Pull Request Aware SZZ Variant} \label{ImprovedDefectLabelingRefs}
Unlike existing variants of the SZZ algorithm which solely rely on data from VCS and issue management systems, our approach utilizes additional information about the software development workflow of a project.
Information contained in pull requests is utilized to propose additions to both steps of the algorithm adding new labeling, filtering and selection mechanisms.
The general premise and structure of SZZ is not changed.
The remainder of this section is composed as follows.
We first describe the data used by our approach, its characteristics and how it is processed to extract needed information.
Thereafter, we describe the improvements, first for finding a fixing commit and second for tracing the inducing commit for a bug ticket.
%%%%%%%%%%%%%%%%%%%%%%%

\subsection{Reconstructing pull-based development data} \label{ReconstructingPRs}

Data from pull-based development includes information about purpose, approach and resolution of a feature.
In the best case scenario each pull request concerns a single feature regardless of whether it is an enhancement, evolution or an attempt to fix a bug.
In this study, only accepted pull requests are considered, as only those related changes were merged and can fix and induce bugs.
Each pull request consists of a title, a description, referenced commits, and collaborative data such as comments and code reviews.
Dependent on the merge strategy, the commits related to a pull request may not be present in the VCS history even if the pull request is accepted.
For instance, when a pull request is accepted and integrated by rebasing the commits, new commits are generated with new hashes.
When a pull request is squashed at merge, a single new commit is created incorporating all change information. 
However, the meta-information (e.g., timestamps, portions of change) and progression of the original commits are shadowed.
We denote commits referenced in pull requests as inner commits. 
We reconstruct the merge strategy from available data and assign inner commits to its VCS equivalents.
In case of squashed or re-based pull requests, we restore the relationship of inner commits to VCS commits by comparing hashes and commit messages.
For squashed pull requests, we extract a list of inner commits mapped to one VCS commit, while otherwise maintaining a one-to-one mapping between commits.
Furthermore, for each pull request the last commit before, first commit after and the resolving commit (the commit with which the pull request is accepted) is obtained.
The processed and reconstructed data is used in both steps of the pull request aware SZZ (PR-SZZ) implementation.

\subsection{Handling duplicate bug information} \label{DuplicateBugsRef}

Our approach considers bug tickets from multiple ITS as well as pull requests that can contain bug information by
describing problem and resolution in one place.
Moreover, projects may use multiple ITS. 
In order to determine if an issue describes a bug, we first check the issue label provided by the systems.
Only if an issue ticket contains a bug label and is already marked resolved, it is considered in this study. 
In accordance with existing work, we assume that all bug labeled items from the CMP and ITS are valid.

The same bug can then be reported in multiple systems with different identifiers.
We automatically join bug tickets recorded in several systems that describe the same issue
and reduce the chance of confusing identifiers of separate systems.
We analyze links between bug tickets and combine all affected instances into one by considering all separate identifiers to point to the same problem.
We argue that this might also improve existing approaches by providing additional issue references for combined duplicate bugs.

\subsection{Matching Bug Fixing Commits} \label{FindingBFCRef}

%problem 1
Existing approaches rely on linking commit message and ticket identifier to match fixing commits to bug tickets. 
This requires a well-formed and strict policy to mention issue identifiers in a certain way so that parsing the issue reference can succeed. 
However, the applied contribution guidelines and commit message structures may vary from project to project.
Thus, the quality depends on the investigated project and the compliance to standards when committing changes.

\begin{figure}
  \centering
  \begin{tabular}{c}
    \includegraphics[width=.95\columnwidth]{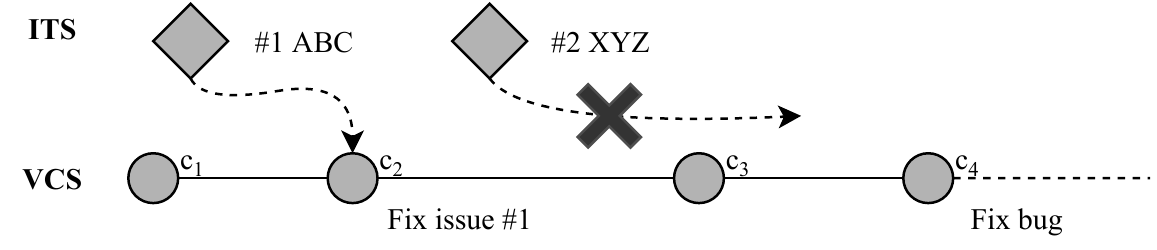} \\
    \footnotesize{(a) Previous process} \\
    \\[.1cm]
    \includegraphics[width=.95\columnwidth]{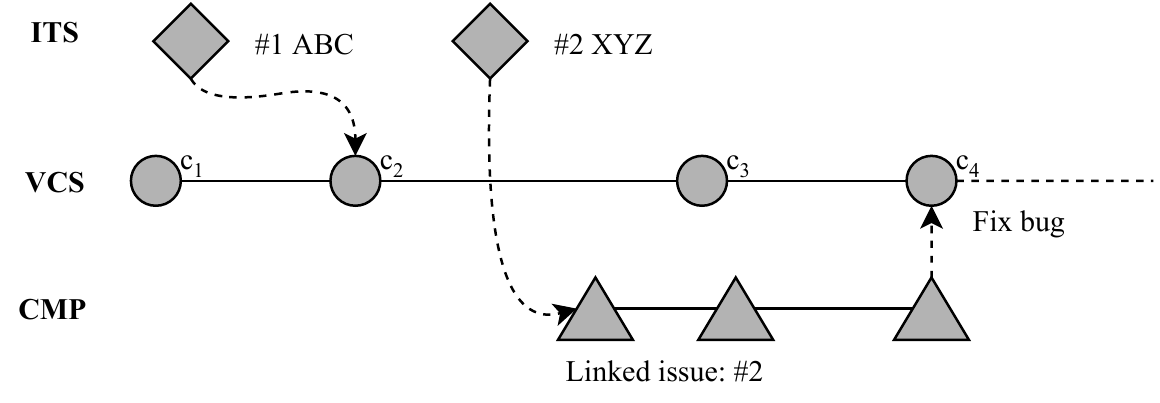} \\
    \footnotesize{(b) Pull requests based approach}
  \end{tabular}
  \caption{Schematic representation for searching fixing commits. Commits in the VCS are shown as circles and denoted with a lowercase c, inner commits of pull requests are shown as triangles. Tickets from the ITS are depicted as squares.}
  \label{fig:fixing}
\end{figure}

%solution
We describe the steps to apply implicitly available data from pull requests to support the first step of SZZ.
Since pull requests are used to incorporate changes into an existing code base, they often include information about resolved issues.
So far, bug fixing commits are retrieved by mapping commits and bug tickets as depicted in Figure \ref{fig:fixing}a.
The mappings rely solely on message parsing and can only find directly mentioned commits (\textit{c\textsubscript{2}}).
If such references are not present, no ticket can be mapped to the fixing commit.
Instead, bug tickets are often linked to pull requests as they are more closely related in the workflow.
Pull requests contain all performed development activities to fix a problem, and the fixing commit can be retrieved from the inner pull request commits (\textit{c\textsubscript{4}}), as abstracted in Figure \ref{fig:fixing}b.
Applying additional data from pull requests, a fixing pull request can be assigned to the issue.
The commit contained in said pull request can be linked directly to the correct fix commit from the VCS history.

%Approach
Our approach starts with extracting all links between pull requests and issue tickets.
Links generated by the domain model of the ITS and CMP are leveraged to get secured links between the systems.
In general, we extract three kinds of integrated links.
%First
First, we utilize the integrated link-section of the ITS ticket view to extract mentions of related items.
For instance, in Jira those links are called "Issue Links", in GitHub they are referred to as "Linked pull requests" or "Linked issues". 
The elements linked here contain a short description and hyperlink to the referenced information. 
%Second
Secondly, we retrieve links created by the platforms based on user interaction and bot integration.
For example, if a GitHub issue is mentioned on a different issue or pull request, the platform creates a mention in the original issue.
%Third
Last, we collect automatically generated links provided by the platforms if a specific syntactic structure is applied in a text.
To illustrate, typing a \textit{\#} followed by a number in GitHub creates a reference to the issue with the specified number.
Apart from these integrated cross-references, we extract links from the title, description, comments, and reviews that mention other pull requests or issues.
A regular expression is applied to obtain these supplementary links from text.
To cope with different ITS, adjusted expressions are crafted for the different systems and applied accordingly.
For GitHub, we use an expression according to the issue linking syntax described above.\footnote[1]{\smaller{ \url{((close[s|d]?|fix[es|ed]?|resolve[s|d]?)+[\s|:]*\#number\D)|(\(\#number\)\D)}}}
In line with results from Herbold et al. \cite{Herbold2019}, for Jira bug tickets the project identifier as well as the issue identifier (e.g., KAFKA-9176\footnote[2]{\smaller{ \url{https://issues.apache.org/jira/browse/KAFKA-9176}}}) is used.\footnote[3]{\smaller{ \url{[\s|:|\[|\(]*(identifier-number)[\]|\)|,|;|:|\b]*}}}

Using the extracted links, a directed graph can be created mapping related elements form the various systems.
Subsequently, we analyze transitive relations and add inferred edges to the mapping.

Searching for fixing commits, PR-SZZ processes all reported bugs and checks if a link to any pull request is present in the constructed mapping.
If exactly one pull request is mapped, it is directly considered as the fixing pull request.
If more than one pull request is linked to a bug, we apply a syntactic and semantic check to retrieve the correct fixing pull request.
We use a confidence level approach similar to the original study \cite{Jaceksliwerski2005}. 
We start with a confidence level of 0 and increment it every time a condition is met.
To check the syntactic validity a link analysis is performed where the confidence level is increased if the pull request has a link to the issue and if the issue has a reference to the pull request.
When applying the semantic checks we evaluate if the assignee of the pull request matches the assignee of the issue and if the creation and closing of the pull request is closest to respective issue events. 
The pull request with the highest confidence is selected as the most probable option to fix the bug.
If no pull request for a given bug ticket is found, we use the approach from the original paper \cite{Jaceksliwerski2005} to map commits directly to the bug ticket using the same adjusted regular expressions described above.

\begin{figure*}[htbp]
\centering
\includegraphics[width=.8\textwidth]{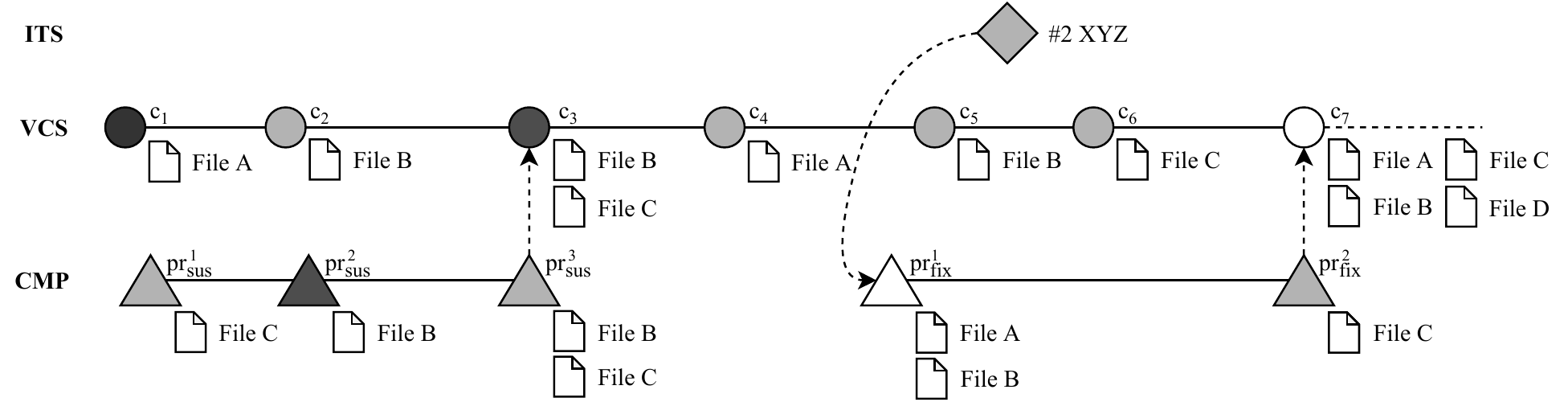}
\caption{A schematic representation for tracing bug inducing commits utilizing pull requests. Commits in the VCS are shown as circles and denoted with a lowercase c\textsubscript{i}, inner commits of pull requests are shown as triangles, denoted as pr\textsuperscript{j}. Tickets from the ITS are depicted as squares. Identified fixing commits are colored white; bug inducing commits are depicted black; all other elements are grayed. Files changed in a commit are shown next to it.}
\label{fig:inducing}
\end{figure*}

Multiple commits can be suspects to fixing a bug, either by selecting a pull request with multiple commits or by mapping multiple commits through commit message parsing.
If more than one commit of the VCS is mapped, we apply a second set of conditions to obtain the correct fixing commit.
Therefore, PR-SZZ applies similar conditions as proposed in the original SZZ on all suspect commits.
The selected commit is marked as the fixing commit for the investigated bug ticket.
A data set of bug tickets with their corresponding fixing commit is created and leveraged in the second step of the algorithm.

\subsection{Tracing Bug Inducing Commits} \label{TracingBICRef}

We build on the approaches described in Section \ref{BackgroundRefs} to trace bug inducing commits. 
We propose extensions to filter changes of the fix commit (\textbf{f\textsubscript{1-3}}) as well as selection mechanisms (\textbf{s\textsubscript{1-3}}) based on pull request information to reduce the number of incorrectly inferred inducing commits.
The fixing commit as well as the suspected inducing commits can be part of developments originated in a pull request.
The pull request for the fixing commit is defined as pr\textsubscript{fix}, the one integrating the suspected bug inducing commit is denoted as pr\textsubscript{sus}.
A schematic representation of VCS, ITS and CMP for an exemplary use case is shown in Figure \ref{fig:inducing}.

For each fixing commit SZZ traces inducing commits that changed the fixed lines.
% problem
In all existing approaches, the bug fixing commit is compared to the previous commit in the VCS history to compute the changes.
However, this may also include lines that are itself part of pr\textsubscript{fix}. 
These changes are most probably not part of the problem but shadow the actual problem solution.
% solution
We calculate the commit difference not with its parent, but with the first parent %commit 
in the VCS history that is not part of pr\textsubscript{fix} (\textbf{f\textsubscript{1}}).
Thereby, all changes contained in pr\textsubscript{fix} prior to the fixing commit are ignored.
In Figure \ref{fig:inducing}, this has no effect since squashed pull requests are always represented by a single commit in the VCS. 
Therefore, commit c\textsubscript{7} is compared to c\textsubscript{6}. 
In other cases this may span several commits.
% problem
As shown in Figure \ref{fig:inducing}, pr\textsubscript{fix} could be squashed so the fixing commit may have several inner commits assigned (pr$^1_{\text{fix}}$, pr$^2_{\text{fix}}$).
%solution 
We obtain all inner commits of pr\textsubscript{fix} and neglect changed files of the selected fixing commit that were not part of any inner commit (\textbf{f\textsubscript{2}}). This may occur when the fixing commit integrates multiple pull requests at once. 
Thus, changes to \textit{File D} of \textit{c\textsubscript{7}} are ignored.
Furthermore, with the same confidence level approach as described for finding bug fixing commits, PR-SZZ tries to select the actual fixing commit from the inner commits (\textbf{f\textsubscript{3}}). 
It neglects all files that were not part of this commit.
Therefore, \textit{File C} in commit c\textsubscript{7} is ignored since the inner fix commit (pr$^1_{\text{fix}}$) does not change the file.
As too large changes tend to generate false-positive inducing commits, a threshold for maximum changed files and lines is applied.
Only the files remaining after applying \textit{f1}-\textit{f3} are processed further.

For each remaining file, the changed lines are filtered.
In accordance with Kim et al. \cite{Kim2006} and Willams and Spacco \cite{Williams2008}, we neglect line changes concerning only whitespaces, imports or comments. 
To detect such changes we build a tool to extract language constructs from code using ANTLR \cite{10.5555/2501720}.
This approach is chosen as it is easily extendable for additional grammars and does not depend on language specific tooling.
Resembling MA-SZZ \cite{Costa}, we ignore meta changes. Assumed that \textit{c\textsubscript{4}} changes only permissions to \textit{File A}, it is ignored and the next commit in the history is blamed.
We consider removed as well as added lines when tracing the bug inducing commits.
Since solely added lines cannot be traced, we resort to considering changes to the whole method body,
as fixes are mostly co-located with their induction \cite{WenMing} (e.g., a missing null check).
We blame the commits that last changed the computed lines.
Each line change is tracked in a line mapping approach to better trace large chunks of changed lines \cite{Williams2008}.
We do not apply refactoring detection tools to be more language agnostic.
The entirety of blamed commits contained in the mappings, form the set of suspicious bug inducing changes.

In order to reduce the list of suspicious inducing commits, we design selection mechanisms based on pull request information.
We extend R-SZZ to not only select the most recent commit before the issue was reported, but also before pr\textsubscript{fix} was created (\textbf{s\textsubscript{1}}). 
In some cases the pull request is provided first and subsequently a matching ticket is created in the ITS.
In the example case, \textit{c\textsubscript{5}} is ignored.
We further propose two new selection mechanisms based on pull requests.
Each pr\textsubscript{sus} for every bug inducing commit is retrieved. 
We obtain links between pr\textsubscript{sus} and pr\textsubscript{fix}. 
If such link exists, we consider the pr\textsubscript{sus} and the associated commits as secured bug inducing commits (\textbf{s\textsubscript{2}}). 
In our example \textit{c\textsubscript{3}}, changing \textit{File B}, would be marked as secured inducing commit if the pull requests are linked.
Defect data sets generated by PR-SZZ are contain on commit, file and method level to enable more fine-grained analyses.
The last selection mechanism (\textbf{s\textsubscript{3}}), subsequently, analyzes pr\textsubscript{sus} in more detail. 
For each changed file and method in a fix commit, the matching inducing file and method is retrieved from the list of suspects.
We select commits that concern the changed lines and ignore changes to different files. 
In the case described in Figure \ref{fig:inducing}, only pr$^2_{\text{sus}}$ and pr$^3_{\text{sus}}$ change \textit{File B}.
However, the blamed lines are only touched by pr$^2_{\text{sus}}$.
Consequently, changes of c\textsubscript{3} are marked as inducing, but only changes in pr$^2_{\text{sus}}$ are considered for the fine-grained data set.

%%%%%%%%%%%%%%%%%%%%%%%
\section{Study Design} \label{EmpiricalStudyRefs}
We evaluate the additions of PR-SZZ and compare the results to existing approaches by conducting experiments for a selection of open-source projects. 
First, we describe the studied projects and their overall characteristics.
Second, the process of data collection and preparation is described.
Finally, we present the evaluation setup for the experiments to address the following research questions:

\begin{itemize}
    \item \textbf{RQ1:} To which degree can pull request data increase the performance of finding bug fixing commits?
    \item \textbf{RQ2:} To which degree can pull request data increase the performance of tracing bug inducing commits?
\end{itemize}

\subsection{Studied Projects} \label{studied_projects}
\begin{table}
\caption{Overview of projects analyzed in this study. Total number of commits, pull requests and tickets is shown alongside number of distinct bugs.}
\label{tab:projects}
\begin{center}
\begin{tabularx}{\columnwidth}{Xrrrr}
\toprule
    Project & \#Commits & \#Pull Requests & \#Tickets & \#Bugs \\
    \hline %% add labeled as bugs per cell?
        angular/angular\textsuperscript{[2]} & 19399 & 11645 & 19282 & 4855 \\ 
        apache/airflow\textsuperscript{[1,2]} & 11042 & 8645 & 7169 & 2938 \\ 
        apache/calcite\textsuperscript{[1]} & 3156 & 1642 & 2910 & 1805 \\ 
        apache/kafka\textsuperscript{[1]} & 7251 & 8036 & 6712 & 3726 \\ 
        apache/pulsar\textsuperscript{[1,2]} & 5790 & 5557 & 1989 & 1750 \\ 
        jenkinsci/jenkins\textsuperscript{[1]} & 10623 & 2812 & 3054 & 2357 \\
\bottomrule
\end{tabularx}
\end{center}
\emph{[1]:} Project uses Jira as ITS \newline
\emph{[2]:} Project uses Github Issues as ITS
\end{table}

In this paper, we investigate pull requests on GitHub as well as bug tickets from Jira and the GitHub issue tracking solution (from here on called GitHub Issues).
We collected data from 20 popular open-source repositories, with a focus on projects from the Apache Software Foundation since these are oftentimes used in other studies concerning SZZ evaluations (e.g., \cite{Fan2019, Herbold2019, Costa}). 
Statistics for all 20 projects can be found in the supplementary material provided alongside this paper.
This study is restricted to projects with at least 1000 stars on GitHub showing community interest, at least 1000 commits to ensure a sufficient development history, a issue-commit ratio of at least 20\% to guarantee ITS usage and a 20\% pull request-commit ratio to ensure the usage of pull requests for integrating changes.
Six projects that meet these criteria are selected as shown in Table \ref{tab:projects}. 
One of the projects solely uses GitHub Issues, three use Jira while two project use a hybrid workflow with both Jira and GitHub Issues configured.

Since our approach relies on a pull-based workflow, the observation period is restricted to a time frame where mostly pull requests are used to integrate changes in the studied projects.
Therefore, for each studied project, we evaluate six years of historical data from January 2015 to January 2021.

\subsection{Study Process} \label{StudyProcessRefs}

In this section, we describe our approach to execute the study.
First, all repositories are cloned and commits from all branches are extracted.
We use the GitHub API to get all pull request information and reconstruct the development activities as described in Section \ref{ReconstructingPRs}. 
For each project, based on the used ITS, we utilize the Jira or GitHub API to get resolved tickets.
Bug labels can be assigned to Jira tickets, GitHub tickets and pull requests.
By analyzing the links between all systems, we find that up to 458 duplicate bugs exist with a mean of 144 duplicates per project.
Hence, we resolve duplicated tickets as described in Section \ref{DuplicateBugsRef} and show the number of distinct bug tickets in Table \ref{tab:projects}.
Similar to existing research, it is assumed that the labels assigned to tickets are correct.
However, it should be noted that this assumption does not always hold according to other studies (e.g., \cite{Herzig2013, Herbold2019}).

We execute PR-SZZ to map bug fixing commits to their corresponding bug ticket as described in Section \ref{FindingBFCRef}.
To evaluate its performance, multiple existing implementations are also executed.
First, we execute OpenSZZ \cite{Lenarduzzi2020}, that implements the original version described in B-SZZ and relies on the same regular expressions as proposed by \'{S}liwersky et al.\cite{Jaceksliwerski2005} for finding the fixing commits.
Second, we re-implemented B-SZZ (B-SZZ*), to adjust the commit message parsing for nowadays established commit message rules. We apply the same regular expressions as described in Section \ref{FindingBFCRef}.
Additionally, another open-source implementation called SZZUnleashed \cite{Borg2019} is executed using the same regular expressions.
All algorithms are adjusted to handle Jira, GitHub Issue and pull request bug descriptions as well as duplicated bug tickets.

Using the bug fixing commits obtained by the first algorithm step, we execute the second step of PR-SZZ to obtain the corresponding bug inducing commits as described in Section \ref{TracingBICRef}.
In order to compare results from PR-SZZ to already existing approaches, we utilize PySZZ \cite{Rosa}, an open-source tool which implements the bug inducing part of B-, AG-, \mbox{MA-}, L-, R-, and RA-SZZ.
RA-SZZ is omitted in our experiments, as the current implementation only supports Java projects.

\subsection{Evaluation Setup} \label{StudyDesignRefs}

With the described study process, we obtain results for mapped bug fixing commits and traced bug inducing commits for all executed SZZ variants.
For both research questions, we evaluate the performance of each executed variant.
Due to the lack of established ground truth for the investigated projects, we evaluate the results first on a quantitative basis 
followed by an evaluation performed on a manually collected and validated data set. 
The performance evaluation is based on precision, recall and F-score.

For the manual evaluation of mapped bug fixing commits (\textit{RQ1}), we select and validate a subset of all bug tickets.
When all algorithms report the same fixing commit for a bug ticket we excluded those from further evaluation since a consensus exists and no alternative result could be extracted by any current implementation. 
Among the remaining bug tickets, we randomly select 50 samples per project and search manually for their corresponding fixing commit.
To identify the correct fixing commit, we use the ticket description on Jira or GitHub, pull request information and the commit message of all commits. 
If we cannot find any fixing commit, for example when the bug ticket was not resolved but closed, we put a null value.
This results in a total of 300 manually validated bug tickets.
When evaluating the performance for PR-SZZ, B-SZZ*, OpenSZZ and SZZUnleashed we collect all identified bug fixing commits for each respective variant, compare it to the correct fixing commits from the validated data set and calculate the performance metrics. 
We report the results in Section \ref{ResultsRefRQ1}.

For the manual evaluation of tracing the correct bug inducing commits (\textit{RQ2}), we again neglect the samples where there is unison among all variants and take a random set of 50 fixing commits for each project.
Furthermore, we remove samples which are not code related (e.g., fixes for documentation) and replace them by additional random samples.
To trace the correct inducing commits manually, we use the change patches from the fixing commits, corresponding pull requests and the git blame feature. We observed that developers often linked to the inducing commit in the discussion of the bug ticket and used them whenever possible.
A single bug can be introduced by multiple commits, therefore we search for the list of inducing commits.
In total, 300 bug fixing commits were randomly selected and manually validated using this approach. All identified inducing commits from each executed SZZ variant are compared to the correct inducing commits in this data set. We compute the performance metrics and analyze the results in Section \ref{ResultsRefRQ2}.

\section{Results} \label{ResultsRef}
This section is organized based on the formulated research questions and discusses them accordingly.
We provide all data sets, necessary source code and detailed evaluation results in our supplementary material.\footnote[4]{\url{https://doi.org/10.6084/m9.figshare.16831492}}

\subsection{RQ1: To which degree can pull request data increase the performance of finding bug fixing commits?} \label{ResultsRefRQ1}

\begin{table}
\caption{Percentage of bug tickets for which a fixing commit was found across all studied projects. Total number of bug tickets in brackets.
}
\label{tab:fixing_commits_percentage}
\begin{center}
\begin{tabularx}{\columnwidth}{lXXXX}
\toprule
    Algorithm & Combined \newline\newline \smaller{(n=17431)} & Pull Requests \newline \smaller{(n=2797)} & GitHub \newline Issues \newline \smaller{(n=5333)} & Jira Issues \newline \newline \smaller{(n=10155)}\\
    \hline
        B-SZZ* & \multicolumn{1}{r}{34.68\%} & \multicolumn{1}{r}{86.56\%}  & \multicolumn{1}{r}{14.06\%} & \multicolumn{1}{r}{35.31\%} \\ 
        OpenSZZ & \multicolumn{1}{r}{16.46\%} & \multicolumn{1}{r}{4.61\%} & \multicolumn{1}{r}{1.84\%} & \multicolumn{1}{r}{27.18\%} \\ 
        SZZUnleashed & \multicolumn{1}{r}{43.34\%} & \multicolumn{1}{r}{65.61\%} & \multicolumn{1}{r}{12.30\%} & \multicolumn{1}{r}{56.18\%} \\ 
        PR-SZZ & \multicolumn{1}{r}{\textbf{61.71\%}} & \multicolumn{1}{r}{\textbf{98.78\%}} & \multicolumn{1}{r}{\textbf{49.43\%}} & \multicolumn{1}{r}{\textbf{61.00\%}} \\ 
\bottomrule
\end{tabularx}
\end{center}
\end{table}

To answer \textit{RQ1}, we first analyze the results of retrieving bug fixing commits on a quantitative basis.
Table \ref{tab:fixing_commits_percentage} contains the percentage of bug tickets with an assigned bug fixing commit across all projects.
PR-SZZ retrieves most fixing commits in all of the investigated projects. 
Since not all fixing commits reference the fixed bug ticket in the commit message, at least 18.37\% of them can be additionally mapped using PR-SZZ. 
We find that for 55.69\% of the bug tickets PR-SZZ would map a fixing commit using links to pull requests, meaning that in 6.02\% of the cases no pull request could be found and PR-SZZ did fall back on searching the commit messages for the corresponding ticket number.
All executed SZZ variants find the same result for 41.04\% of all bug tickets, with no variant being able to retrieve a fixing commit for 36.05\%. 
We see that there are still many bug tickets with no assigned fixing commit. 
However, this is not necessarily wrong since not all resolved bug tickets are fixed by a commit but may be closed due to irrelevance or because they were fixed outside of the project scope. 

PR-SZZ finds a fixing commit for a bug ticket in most cases compared to the other algorithms regardless of where the bug ticket originates.
When comparing the results for the different ITS, we observe some variations in the gain PR-SZZ provides compared to existing solutions.
It becomes apparent that PR-SZZ outperforms alternatives when GitHub Issues is used. It finds a fixing commit for another 35.37\% of bug tickets compared to the most finding alternative (i.e., B-SZZ*). When looking at bug tickets from Jira the difference to the best alternative (i.e., SZZUnleashed) is smaller with additionally found fixing commits for 4.82\% of bug tickets.
On the project level, this is a recurring theme. 
The fact that projects using GitHub Issues can map extensively more fixing commits to bug tickets may be due to the better integration of tickets and pull requests in the platforms ecosystem. 
Using Jira it is evident that links between bug tickets and pull requests are present less often, however ticket identifiers are more often added to the commit message by the developers.
It seems that the automation in GitHub reduces manual issue tracking effort for the developers. Therefore, with more open-source projects using GitHub Issues, new data such as pull requests is needed to retrieve fixing commits.
An evaluation for each project separately can be found in the supplementary material provided with this paper.

\begin{table}
\caption{Performance evaluation of finding bug fixing commits for the executed SZZ variants averaged over all projects.}
\label{tab:fixing_commits_performance}
\begin{center}
\begin{tabularx}{\columnwidth}{Xrrrr}
\toprule
    %Perspective & (1) & \multicolumn{3}{r}{(2)} \\
    Algorithm & Precision &  Recall & F-score \\
    \hline
        B-SZZ* &  0.74 &   0.24 &  0.35 \\
        OpenSZZ & 0.34 &   0.12 &  0.16 \\
        SZZUnleashed & 0.75 &   0.26 &  0.34 \\
        PR-SZZ & \textbf{0.79} &  \textbf{0.72} &   \textbf{0.75} \\
\bottomrule
\end{tabularx}
\end{center}
\end{table}

That said, the number of found fixing commits does not reflect on the quality of the inferred mappings. Therefore, we continue with evaluating the performance of PR-SZZ compared to existing open-source implementations on a manually validated subset of bug tickets, as described in Section \ref{StudyDesignRefs}. 
The results are shown in Table \ref{tab:fixing_commits_performance}.
With the correct fixing commit found in 71.67\% of the sampled bug tickets, PR-SZZ is the most reliable algorithm among the variants.
On average, PR-SZZ increases precision as well as recall for the manually validated data sets. 
For the individual projects precision is increased for three, while recall is highest for all of the projects.
The especially high recall on all projects shows that additional fixing commits can be correctly mapped to the sampled bug tickets and the high precision states that the found bug fixing commits are mainly correct resulting in an higher F-score on average and for every distinct project under investigation.
Looking at the performance in regard to the different ITS no apparent difference can be noticed.
We conclude that the quality of the identified bug fixing commits is not determined by the use of ITS for any algorithm.

\vspace{.15cm}
\noindent\fbox{\parbox{.97\columnwidth}{\textit{
Using PR-SZZ, 18\% of bug tickets can be additionally mapped to a fixing commit.
On average, precision as well as recall is increased, resulting in an F-score of 0.75
compared to 0.35 for the best performing existing alternative.
We conclude that using pull request data contributes to the completeness and performance of the first step of the SZZ algorithm.
}}}
\vspace{.1cm}

%%%%%%%%%%%%%%%%%%%%%%%%%%%%%%%%%%%%%%%%%%%%%%%%%%%%%%%%%%%%%%%%%
%%%%%%%%%%%%%%%%%%%%%%%%%%%%%%%%%%%%%%%%%%%%%%%%%%%%%%%%%%%%%%%%%
%%%%%%%%%%%%%%%%%%%%%%%%%%%%%%%%%%%%%%%%%%%%%%%%%%%%%%%%%%%%%%%%%

\subsection{RQ2: To which degree can pull request data increase the performance of tracing bug inducing commits?} \label{ResultsRefRQ2}
The second part of SZZ uses mapped fixing commits as input and traces the inducing code. 
We use the data set created by the first step of PR-SZZ from above as input for this evaluation.
In this data set the fixing commit has a reference to a pull request in 91.37\% of cases. 
Inducing commits are found for 63.76\% of all bug tickets for PR-SZZ and using existing approaches for between 54.08\% (e.g., MA-SZZ) and 67.97\% (e.g., B-SZZ) of the cases.

Hereafter, we evaluate the number of cases where our proposed filter (\textbf{f1-f3}) and selection (\textbf{s1-s3}) mechanisms from Section \ref{ImprovedDefectLabelingRefs} are applicable for the investigated projects.
Comparing a fixing commit to its direct parent inside the pull request instead of the latest not feature related commit (\textbf{f1}), the traced inducing commits can be part of the bug fix.
In case of B-SZZ, results show that a total of 205 selected inducing commits are part of the fixing pull request. The inducing commit for a bug ticket, however, cannot be created within the pull request to fix the actual bug.
Files from the fixing commit that are not modified by the fixing pull request are found in 83 cases (\textbf{f2}).
For squashed fixing commits, we find that an inner commit can be selected as real fixing commit in 2571 cases (\textbf{f3}).
In total, the number of traced files can be reduced in 718 cases.
The fixing pull request is created before the bug ticket (\textbf{s1}) in 467 cases, which enables PR-SZZ to ignore suspicious inducing commits that can not be responsible for the introduction of the bug.
Additionally, a suspect inducing commit is mentioned in the fixing pull requests (\textbf{s2}) a total of 237 times, providing additional context for the selection of a inducing commit.
The selected suspect is squashed in 1069 cases, allowing further filtering and selection for file and method level data sets (\textbf{s3}).
This shows that the proposed additions can have an impact on the results of SZZ and may improve the performance in terms of correctly traced and selected bug inducing commits for the corresponding cases.

PR-SZZ reports an average of 4.19 inducing suspects ranging from 1 to 108 inducing commits per bug fixing commit.
Existing approaches do report more inducing commits (maximum 463 for MA-SZZ), with an average between 1.61 (e.g., AG-SZZ) and 4.34 (e.g., B-SZZ) per bug fix.
However, only 14 of the 300 manually validated bug fixing commits had multiple inducing commits (maximum 4 suspects).
With many suspects found for a single bug fix many false-positives are introduced in the result set.
We argue that it is important to reduce the number of false-positives since the data sets are used as input for research on bug characteristics and defect prediction, where wrongly labeled bug inducing commits may have implications.
With the selection of one commit per bug fix the number of wrongly labeled inducing commits can be reduced.
When selecting a single commit the results of all selecting variants (L-, R- and PR-SZZ select) do overlap.
On average, 28.56\% (min: Kafka with 20.85\%; max: Airflow with 36.42\%) of selected inducing commits from PR-SZZ are also selected by either L- or R-SZZ. This shows that PR-SZZ does select new inducing commits for the majority of cases.

\begin{table}
\setlength\abovecaptionskip{-0.4\baselineskip}
\setlength{\textfloatsep}{5pt}
\caption{Performance evaluation of tracing bug inducing commits. The results for PR-SZZ are reported for all suspects (PR-SZZ) and after the selection mechanism is applied (PR-SZZ selected).}
\label{tab:inducing_results}
\begin{center}
\begin{tabularx}{\columnwidth}{Xcrrr}
\toprule
Algorithm & Selection Applied? & Precision & Recall & F-score \\
\midrule
B-SZZ       &  \xmark      &  0.34 &   0.62 & \textbf{0.44} \\
AG-SZZ       & \xmark        &  0.26 &   0.21 & 0.23 \\
MA-SZZ      & \xmark       &  0.19 &   0.26 & 0.22 \\
PR-SZZ      & \xmark      &  0.30 &   \textbf{0.68} & 0.41 \\
L-SZZ       & \cmark       &  0.32 &   0.19 & 0.23 \\
R-SZZ       & \cmark     &  0.29 &   0.17 & 0.22 \\
PR-SZZ selected     & \cmark       &      \textbf{0.50} & 0.39 & \textbf{0.44} \\
\bottomrule
\end{tabularx}
\end{center}
\end{table}

In order to evaluate the added value of proposed adjustments, a manual evaluation is conducted as described in Section \ref{StudyDesignRefs}. 
The results are shown in Table \ref{tab:inducing_results}.
On average, B-SZZ and PR-SZZ report a significantly higher F-score compared to all other variants.
Looking at each single project, a more diverse distribution becomes apparent. 
B-SZZ has the higher precision in 5 projects, with PR-SZZ in the other project (mean difference of 0.03).
While PR-SZZ has the higher recall on average and in 5 projects, with B-SZZ in the other project (mean difference of 0.07). 
Please refer to the supplementary material for a in depth comparison.
This states that PR-SZZ can find more correct suspects than all other variants.
However, it also flags more commits that are indeed not inducing bugs into the code base.
With the high number of suspects for PR-SZZ and B-SZZ, described in the quantitative analysis, they show by far the highest and second highest recall value. 
The higher recall value, however comes with a trade-off in precision. 
Looking at the executed selecting variants, PR-SZZ reports the highest precision among all variants, outperforming the next best approach (L-SZZ) by 18.41 percentage points. The F-score of 0.44 is comparable to the one reported by B-SZZ showing clearly the trade-off between precision and recall.
This also shows that the proposed filtering and selection mechanisms work, lead to more correct suspects, selecting the correct ones more often. 
Looking at each project, we see that the PR-SZZ selection mechanism has the highest precision value for all projects. Moreover, it states the highest F-score in three of the projects.

We conclude that PR-SZZ needs specific workflow attributes to be fully applicable.
For instance, in projects with a primarily squashed workflow, inner commits of pull requests can be used to filter the list of suspicious files, which in non-squashed workflows mitigates the effect size. 
Applying PR-SZZ with its selection mechanisms, the precision of found inducing commits on our manual evaluation data set is increased. 
As already mentioned, this is especially important as data sets created by SZZ are used in other research.

\vspace{.15cm}
\noindent\fbox{\parbox{.97\columnwidth}{\textit{
Evaluating the proposed filtering and selection mechanisms the performance of existing SZZ variants is highly dependant on the project characteristics.
If selection mechanisms are applied PR-SZZ reports the highest precision for all investigated projects, outperforming all other variants by on average 16 percentage points, leading to higher quality data sets.
}}}
\vspace{.1cm}

\section{Discussion} \label{Discussion}

In this section, we discuss characteristics of both steps of the algorithm that influence the performance of SZZ in general and PR-SZZ in specific. Especially, the influence on the investigated projects that vary in size and development workflow is discussed.
Variations in performance are evident in both steps of the algorithm depending on the project and applied SZZ variant.
Analyzing the merge strategies of all projects, we find considerable differences.
Half of the projects use a preliminary rebase-based workflow, while the others use a merge-based workflow.
Furthermore, 3 projects use an predominately squashed integration mechanism.
Exemplary, this leads to 4823 out of 8036 pull requests with shadowed commits for Apache Kafka. 
With different integration strategies, SZZ may compute results of varying quality.

\subsection{Finding Bug Fixing Commits}
An determining factor for the performance of finding the bug fixing commit is the quality of links between tickets and commits for existing SZZ variants and for PR-SZZ also between tickets and pull request.
On average, 51.96\% of bug tickets are linked to a pull request with differences between the projects (min: 35.36\%, max: 64.56\%).
Furthermore, on average, only 7.26\% (min 0.00\%; max 17.71\%) of pull requests were labeled as bugs.
With existing variants relying on matching commit message with bug tickets, additional attention. As shown by the results of OpenSZZ the initially suggested regular expressions are not applicable with nowadays commit message rules.
Therefore, the matching regular expression for bug ticket references in the commit message needs to be adjusted to the project workflow.
With the implicit link used in PR-SZZ, no string matching is needed, which mitigates this problem.
In cases where the issue identifier is not referenced in the commit message, information from pull requests is needed to find the fixing commit.

In the VCS, multiple commits can reference the same bug ticket. 
Determining the correct commit from multiple \mbox{fix-,} \mbox{test-,} or refactoring-commits is not possible for existing SZZ variants since no additional information is used.
Projects that squash commits on integration, so that a branch is only represented as one commit in the VCS history, resolve this problem for the first step of the algorithm. Nonetheless, it creates a new problem for tracing the inducing commit, since the squashed commit contains multiple inner changes, while most of the times only one of the inner changes fixes the problem. 
PR-SZZ handles these special cases by using the linked pull request and selects an inner commit from the gained context.

\subsection{Tracing Bug Inducing Commits}
Tracing inducing commits, the project characteristics and especially the merge strategy influence the results of SZZ and the applicability of PR-SZZ. 
The investigated projects use different merge strategies as well as vary in size and programming language.
Therefore, the applicability of filtering and selection mechanisms varies between projects.
For instance, \textit{f3} can be applied in many instances of the Apache Airflow project but has only little effect on projects such as Jenkins. 

Many validated bug fixes are partial fixes (182 out of 300) with on average 2.47 files in the fixing commit and the validated inducing commits are often partially inducing (275 out of 300) with on average 102 files in the inducing commits, with a small fraction of these files fixing or introducing the problem. With PR-SZZ we apply filter for inner file changes and create data sets on the file and method level to reduce the error rate inside the selected commits. An evaluation of these data sets is contained in the supplementary material.

Interestingly, B-SZZ was the best baseline approach for the investigated projects, with comparable results to PR-SZZ. We see that, existing extensions to SZZ did not perform better in most instances. Most of the additions proposed by AG- or MA-SZZ are only applicable in special cases and do not necessarily increase the performance on the validated data set. We find that oftentimes the performance is worse.
This is in line with the results of Rosa et al. \cite{Rosa}. 
Although B-SZZ performs well in terms of F-score in our study, it reports many false-positives.
A high number of false-positives introduces noise in the data set, which may poses a threat to validity for studies that use the data sets. For example applying low quality data sets in just-in-time defect prediction can influence the performance of such models \cite{8765743}.
However, the SZZ variant used to create these data sets should be selected based on the specific use case.
If it is more important to obtain all possible inducing commits, selection mechanisms can be ignored leading to a high recall. On average, PR-SZZ has the highest recall in our evaluation (+0.06 to B-SZZ). If high quality is more important, the mechanisms proposed in this paper increase precision compared to existing variants (+0.16 to B-SZZ). 

%%%%%%%%%%%%%%%%%%%%%%%
\section{Related Work} \label{RelatedWorkRefs}

Improvements and related approaches to the SZZ algorithm are described in detail in Section \ref{BackgroundRefs} of this study.
In this section, we distinguish ourselves from existing literature.

Prior studies concentrate on one issue management tool when evaluating their approach. 
For instance, Jira \cite{Borg2019}, Bugzilla \cite{Jaceksliwerski2005} or GitHub Issues \cite{Sahal2018} is used.
We show results for different project workflows and ITS and analyze 
their influence on SZZ implementations.
Sahal et al.'s \cite{Sahal2018} SZZ enhancement tracks solely added lines by extending the blamed change to the surrounding code block in Java. 
We extend their approach to other languages and to a broader context. 

In order to identify bug fixing commits, other techniques using heuristics on bug reports and commits in cases where no explicit link is given were proposed in the literature \cite{relink, mlink}. 
However, they do also first apply a search for secured direct links. 
We propose using data from pull requests to uncover new direct links between bug reports and their fixes. For bug reports without a fixing commit after applying our approach, these heuristics may be a good fallback option.

In order to make SZZ comparable multiple studies focus on evaluating the results. 
Da Costa et al. \cite{Costa} propose several universal evaluation criteria. 
We do not use their framework since we focus on benefits from pull requests in general.
Rosa et al. \cite{Rosa} craft an oracle using natural language processing on the commit message and infer mentions to inducing commits. 
Due to this selection process it is not used in our study.

To the best of our knowledge, no existing SZZ variant uses data apart from the VCS history and bug tickets. 
This has foremost historical reasons where proposed enhancements concentrate on code characteristics to boost the algorithms performance.
Our approach uses pull requests to improve the SZZ algorithm.
Regarding pull requests, data generated by pull-based workflows is already used by researchers to support decision making (e.g., \cite{Ye2021, Zhao2019}) and to study pull request characteristics (e.g., \cite{Lenarduzzi2021, Nadri2020, Gousios2013}).
For example, they find that the acceptance of pull requests is not directly influenced by the quality of the proposed code \cite{Lenarduzzi2021}.
Furthermore, Gousios \cite{Gousios2013} found that squashed commits are hidden in VCS.
We apply their findings to propose enhancements to SZZ.
Lastly, Huq et al. \cite{Huq2019} analyse pull requests in the context of inducing commits.
They perform a sentiment analysis on pull request artifacts and find that bug inducing changes show a more negative sentiment compared to regular commits. 
In future our approach may be extended to apply and evaluate their findings.

\section{Threats to Validity} \label{ThreatsToValidityRefs}

\subsection{Internal Validity}

Since no ground truth is available for the investigated projects, we construct the data sets for evaluating our approach manually. 
The samples were labeled by one of the authors following a strict process. However, it is still possible that bias was introduced or samples were wrongly labeled.
We cross-referenced our findings with related literature to verify plausibility.
We assume that all bug labeled items from the collaborative and issue management platforms are valid. 
While manually validating bug tickets, we found that many of them do not describe bugs but rather improvements. 
Furthermore, bug tickets are often concerning changes in documentation or configuration files, with no real commit inducing the problem. 
We circumvent this in the manual evaluation by neglecting samples that were no bug description.
Last, the manual evaluation is based on a random subset of bug fixes and inducing commits.
Thereby, not all data is validated. 
We release all data for other researchers to reproduce our results.

\subsection{External Validity}                  
We selected 6 projects from GitHub to perform our evaluation of different background, languages and with different ITS.
Nonetheless, we cannot be sure that our results can be generalized, as we already report differences in the performance of PR-SSZ with regard to project setups.
We predominantly used Apache Foundation projects to mitigate the problem.
In the future, we plan to evaluate PR-SZZ on a bigger and more diverse set of projects.

%%%%%%%%%%%%%%%%%%%%%%%
\section{Conclusion and future work} \label{ConclusionRefs}

The SZZ algorithm is the standard way to retrieve defect data sets from the version control system using the issue tracking system. 
However, not all bug fixing commits for all bug tickets can be found, while large fixing commits produce a large list of potentially bug inducing commits. 
Given SZZ forms the basis for many empirical software engineering studies, we aim to increase its completeness, performance and granularity.
We propose additions to the SZZ algorithm by using pull-based data as they describe development activities in more detail.
We apply the pull-request-aware SZZ (PR-SZZ) to open-source repositories and found that our additions can support the search for bug fixing and inducing commits.

When searching the \textit{bug fixing commit}, existing SZZ variants report high precision but low recall, meaning that many true positive bug fixing commits are missed. Due to links between bug tickets and pull requests, PR-SZZ is able to find additional bug fixing commits. We report increased precision and recall, elevating the F-score on average by 40 percentage points.
When tracing the \textit{bug inducing commits}, overall low precision values are obtained by all existing variants.
With SZZ building the foundation for subsequent research, precision should be considered more valuable than recall and be the focus of future improvements. 
Applying our proposed mechanisms, the highest precision is observed for all projects with a gain of on average 16 percentage points.  

In order to enable further research, we publish data sets for 6 open-source projects on commit, file and method level.
Future work should focus on the in depth analysis of project environments and workflows to investigate the characteristics determining the performance of SZZ.

\section*{Acknowledgment}
This study is based upon work supported by the \textit{Bavarian Ministry of Economic Affairs, Regional Development and Energy} under the \textit{Center for Code Excellence} project.

\bibliographystyle{IEEEtran}
\bibliography{IEEEabrv,main}
\end{document}